# Holo-Block Chain: A Hybrid Approach for Secured IoT Healthcare Ecosystem

Asad Aftab[1], Chrysostomos Chrysostomou[2], Hassaan Khaliq Qureshi[3], and Semeen Rehman[1]

[1]Technische Universität Wien (TU Wien), Austria
[2]Frederick University, Nicosia, Cyprus.
[3]National University of Sciences and Technology (NUST), Islamabad, Pakistan.
[1](asad.aftab, semeen.rehman)@tuwien.ac.at, [2]ch.chrysostomou@frederick.ac.cy, [3]hassaan.khaliq@seecs.edu.pk

*Abstract*—The Internet-of-Things (IoT) is an imminent and corporal technology that enables the connectivity of smart physical devices with virtual objects contriving in distinct platforms with the help of the internet. The IoT is under massive experimentation to operate in a distributed manner, making it favorable to be utilized in the healthcare ecosystem. However, under the IoT healthcare ecosystem (IoT-HS), the nodes of the IoT networks are unveiled to an aberrant level of security threats. Regulating an adequate volume of sensitive and personal data, IoT-HS undergoes various security challenges for which a distributed mechanism to address such concerns plays a vital role. Although Blockchain, having a distributed ledger, is integral to solving security concerns in IoT-HSs, it undergoes major problems, including massive storage and computational requirements. Also, Holochain, which has low computational and memory requirements, lacks authentication distribution availability. Therefore, this paper proposes a hybrid Holochain and Blockchain-based privacy perseverance and security framework for IoT-HSs that combines the benefits Holochain and Blockchain provide, overcoming the computational, memory, and authentication challenges. This framework is more suited for IoT scenarios where resource needs to be optimally utilized. Comprehensive security and performance analysis is conducted to demonstrate the suitability and effectiveness of the proposed hybrid security approach for IoT-HSs in contrast to the Blockchain-only or Holochain-only based approaches.

*Index Terms*—Holochain, Blockchain, Internet of Things (IoT), IoT Healthcare Ecosystem, Security Attacks

## I. Introduction

In recent times, the Internet of Things (IoT) has evolved a lot in terms of its usage in the heterogeneous ecosystem due to its convenience, timeliness, inclusiveness, integration, scalability, and interoperability. Thus, with the enhancement of heterogeneous technologies, the IoT is rapidly footprinting all facets of our lives, including agriculture, remote healthcare, navigation, industrial automation, intelligent home automation, etc. In particular, the IoT introduction in the healthcare sector applications is beneficial since it connects all the nodes to enable universal availability of remote healthcare for everyone irrespective of their physical locality [1], [2].

In the IoT healthcare ecosystem (IoT-HS), several medical devices are utilized as remote monitoring systems for far-flung patients to provide better medical treatment [3]. Such networks are designed to safeguard patient safety and be intimate about critical conditions of patients' health in case of any emergency [4]. For this reason, small sensor devices implanted on a patient's body or devices carried along by a patient can transmit its data packets, composed of critical and non-critical health data, to the nearby smart server or transmitting entity to forward data to servers.

At the same time, IoT-HS connectivity over wireless channels has arisen critical security vulnerabilities. IoT-HS security concerns have been investigated in recent years, like proposing an authentication mechanism with the help of bi-linear pairing and trusted authority [5], private Blockchain solution [6], and cloud server Blockchain mechanism [7].

Blockchain [8], a distributed, decentralized, and ledger-based mechanism provides a consensus-based approach. Consensus is carried out by Smart Contract (SC), which describes rules for every node to follow in a decentralized transaction. Therefore, it provides privacy, where the data are encrypted and managed in such a way that a ledger describes its authenticity [9]–[14]. Pandi Vijayakumar et. al proposed an efficient and secure anonymous authentication mechanism with location privacy in Wireless Body Area Network (WBAN) [5]. Also, in [6], studies have been carried out to devise a Blockchain framework for data sharing using two private Blockchains: registration of sensor nodes data and the other for super-mode and psychological data of patients.

Similarly, a security mechanism proposed in [7] is based on cloud server blockchain architecture, where WBAN data encryption is done with ciphertext-policy attribute-based encryption (CPABE). Also, in [15], a tier-based end-to-end architecture is proposed to control multiple blockchains. These blockchains ensured both security and privacy with greater scalability features. Moreover, a distributed patient health record (PHR) called Omni PHR is proposed in [16], ensuring both elasticity and scalability by bringing in blockchain technology to minimize the average response time [17].

Under the Blockchain network, every node has a copy of the ledger, and if any ambiguity is found, nodes can raise it. However, since a copy on every node's transaction is created, this causes an increase in storage. Thus, a more efficient and rapid distributed strategy for a system can be utilized to improve the overall security level of IoT while considering the storage memory and computational constraints.

Therefore, Holochain, which is the beginning of a new era for an open-source, decentralized network without building bulky data storage and data exchanges like blockchain [18], can be utilized. Holochain brings the concept of combining two techniques: Distributed Hash Table (DHT) and Hash Chain (HC). The DHT is centered on propagating data in the network, while HCs are developed to sustain data integrity [19]. Therefore, Holochain is considered a better approach compared to Blockchain while catering to the security of IoT-HS [20]. Specifically, in [20], a significant improvement is shown in reducing time complexity, computational power, and memory usage. But, in this strategy, the random selection of transaction approval authorities may poison the IoT-HS network by selecting the attacker to be part of the network. Therefore, a more effective load-balancing technique needs to be devised.

This paper proposes an efficient and robust framework for mitigation measures for several security threats to IoT-HS utilizing a hybrid approach, combining the benefits Holochain and Blockchain offers. In particular, Blockchain, being a distributed ledger, requires public consensus that helps in the authentication process to be decentralized, whereas Holochain caters to data integrity as every data packet can be part of the DHT, and therefore cannot be altered during the transmission, which in case of blockchain will increase memory and computational overheads. The proposed hybrid framework aims to enhance coverage, scalability, storage, robustness, data transmission, and routing, offering low complexity and a secure ecosystem. Our contributions to this paper are as follows:

- A hybrid security framework for IoT-HS is proposed, combining the advantages provided by Holochain and Blockchain.
- An extensive performance analysis is provided over a number of parameters that includes the complexity of the system, data storage consumption, average network latency, average ratio of mean network throughput, and average route selection time consumption.
- Advance and distributed mitigation techniques for security threats are proposed.

The paper is organized as follows. Section II presents the details of the proposed hybrid framework in the IoT-HS. Section III provides the performance analysis of the proposed framework in contrast to its counterparts. Finally, Section IV provides the conclusions and future work.

## II. Holo-Block Chain Framework in IoT-HS

In this section, we propose a hybrid Holochain and Blockchain (Holo-Block chain) based privacy perseverance and security framework in IoT-HS for a highly sensitive network to utilize low memory and computation in a distributed manner. Fig. 1 shows the proposed hybrid

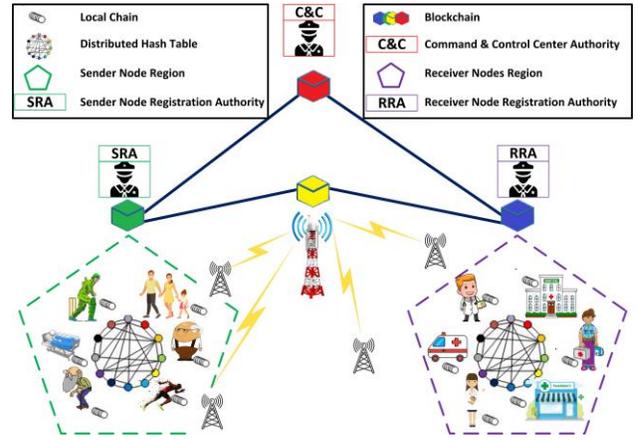

Fig. 1: System Model

IoT-HS framework that includes all entities involved. In particular, the role of each entity is described as follows:

- *Sender Node (SN):* There are several IoT devices that collect, accumulate and forward data without any computation toward the computational end of the network. There are sensors embodied on a patient, athlete, or any entity that requires monitoring and has a short range of communication, thus can propagate their data to a Personal Device (PD). A PD, a handheld device with a larger network coverage area, can thus transfer data to the Base station (BS) to propagate it along the network to the receiver node.
- *Receiver Node (RN):* This entity of the network is responsible for receiving, process, and then taking decisions. These may include the devices with Doctors, pharmacies, paramedics, or any other responsible for monitoring purposes.
- *Holochain:* It is responsible for accumulating the data in a local chain and provides integrity and privacy through DHT. Moreover, the data is validated in the form of an HC.
- *Blockchain:* It is responsible for maintaining a ledger for every new registration and helping mitigate any malicious node interacting with the network. Also, it helps in securing routing paths. All these are executed using SC.

### A. System Model

The formulation of a simple generalized model of a distributed system is given by [21]. $M$ is considered a set of nodes (as denoted in Table I) where $M = \{n_1, n_2, n_3, ....., n_n\}$, deployed in a geographical region with each having location $Loc(n_i)$ where $n_i \in M$. In the case of holochain [21], these $M$ nodes have a state $Z$ associated to it. For each $n^{th}$ node the state $Z_n$ can be a set where $Z_n = \{\sigma_1, \sigma_2, \sigma_3, .....\}$. Therefore, $\sigma_i$ is assumed as $\forall \sigma_i \in SN_n$, where $SN_n$ refers to the sender node, and $\sigma_i = \{\chi_i, D_i\}$, where $\chi_i$ is the set of HC elements associated to each $n_{th}$ node, while $D_i$ is a set of non-HC

TABLE I: Symbol Notation

| Variables-Symbols | Description |
|---|---|
| $M$ | Set of Nodes |
| $Z$ | State of a Node |
| $\chi$ | Hash Chain |
| $D$ | Set of non-hash data elements |
| $H$ | Hash Function |
| $\tau$ | State Transition Function |
| $h$ | Header |
| $tD$ | Transaction Data |
| $\vee(tD, v)$ | Validation Function |
| $St$ | Stimulus Function |
| $P(x)$ | Processing Function |
| $C$ | Channel |
| $c$ | Confidence Complexity |
| $m$ | Total Number of Nodes |
| $Loc$ | Location of PD |
| $En$ | Energy of sensors attached to a PD |
| $S_n$ | Sender Node |
| $\gamma$ | Number of Transactions per node |

data elements. The state transition function of each node is given by [21]

$$\tau(\sigma_i, tD) = (\tau_\chi(\chi_i, t), \tau(D_i, t)) \quad (1)$$

where,

$$\tau_\chi(\chi_i, t) = \chi_{i+1} \quad (2)$$

$$\tau(D_i, t) = D_{i+1} \quad (3)$$

Here,

$$\chi_{i+1} = \{x_1, x_2, ...., x_i, x_{i+1}, ....\} \quad (4)$$

$$D_{i+1} = \{d_1, d_2, ...., d_i, d_{i+1}, ....\} \quad (5)$$

Thus, every $\chi_i, D_i$ set can be represented as given in (4) and (5), respectively. These HC elements are constructed by transforming a transaction data $tD$ by adding a header $h$ to it that is estimated as given by [21]

$$h = \{H(t), y\} \quad (6)$$

where,

$$y = \{H(x_j) | j < i\} \quad (7)$$

Moreover, a stimulus input is required when a transaction needs to float over a network channel $C$. Therefore, the stimulus function $St$ is dependent on $tD$. Thereafter, it needs to be validated whenever $tD$ is generated from a node in a network. To this end, a validation function $\vee$ dependent on $tD$, along with extra bits $v$, is used to validate the $tD$.

Lastly, the processing function $P(x)$, creates the $tD$ and triggers the functions $\vee(tD, v)$ and $\tau$.

**Assumption:** A single type of $tD$ is transferred by a node in the IoT-HS of 256 bytes of data and 12 bytes of additional header. The header of each packet comprises the following fields:

- *Identification*: This 1-byte field provides the unique identification number associated with each distinct type of data.
- *Total Length*: This 1-byte field is associated with the size of the packet length, including the header size of the data generated from a node. It ranges from 12 bytes to 256 bytes.
- *Priority*: This 1-byte field exhibits the urgency of the packet transferred. The priority varies from $k$ (lowest priority) to $l$ (highest priority), where $k, l \in N$.
- *Time to Live*: This 1-byte field helps packets to no longer stay in the network in order to avoid network congestion.
- *Source Address*: The address of the source node generating the data.
- *Destination Address*: The address of the destination node where data is to be delivered.

B. Hybrid Holo-Block Chain Framework

In this sub-section, we present the details of the proposed hybrid Holo-Block chain framework that provides privacy perseverance and security for IoT-HSs, combining the benefits Holochain and Blockchain provide and thus overcoming the computational and memory challenges, along with authentication mechanism challenges. In particular, we combine the advantages provided by the Blockchain in the authentication process and those of Holochain for data integrity. In the following, we provide the details of the utilization of the Blockchain and Holochain technologies under the proposed framework.

*1) **Blockchain Network Definition***: Every entity of the IoT-HS is part of the Blockchain and is registered according to the structure of the hybrid Holo-Block chain framework (*HB*), which has the following attributes:

- *ID*: It refers to the unique identification number associated with each IoT device in IoT-HS.
- *Type*: It refers to the type of IoT device in the IoT-HS, including sensors, handheld device/PD, monitoring devices, transfer gateways, etc.
- *Operational Cost*: It is the cost with which each type of IoT device is in operation. It refers to the time a transaction takes while being distributed on the network.
- *Storage Space*: The total amount of storage space the data packet would take on the storage device.
- *Start Time of Operation*: The start time of operation is noted for every IoT device, so the device's energy consumption starts to be considered.
- *End Time of Operation*: This is needed so as to take some action to ensure the system operation is continued, like battery replacement and/or energy harvesting when the residual energy is too low.
- *Geo-Location*: Every IoT device in the IoT-HS is assumed to have a Geo Positioning System (GPS) to provide its geo-location accordingly.
- *Association*: Each IoT device is associated with one authority.

Fig. 2 represents the actual data flow in the proposed framework and illustrates the authentication process carried out by Blockchain to maintain an upper layer of authentication for registered nodes, thus decreasing network congestion. For this purpose, a private SC with three authorities' signatures is deployed. These three authorities include the Sender Node Registration Authority (SRA), the Receiver Node Registration Authority (RRA), and the Command and Control center (C&C). For any new registration to the system, a two-tier signature method

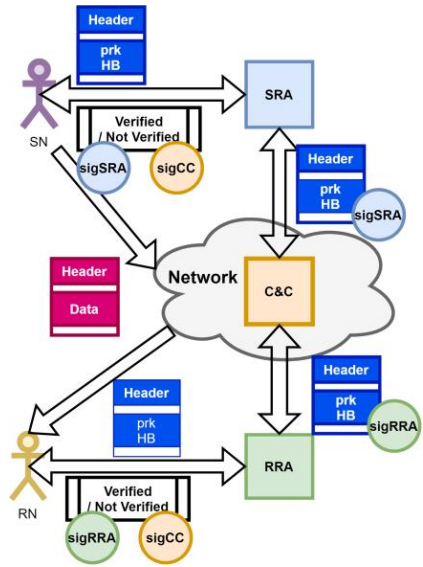

Fig. 2: Data Flow of HB framework

(one from SRA or to which the IoT device is associated, and one from the C&C center) is compulsory.

**Sender Node Registration:** All of the Patients in the IoT-HS, being SNs, have PD and need first to generate private ($prk_{SN}$) and public ($pbk_{SN}$) keys. These IoT devices are registered to the Blockchain C&C center by the signatory authorities SRA and C&C by providing all the values to construct *HB*, $pbk_{SN}$, and digital certificate ($DCert_{SN}$) to the SRA, namely,

$SN_m \rightarrow SRA : requests_{SN_m} = \{HB||pbk_{SN}||DCert_{SN}\}$

As soon as the SRA receives the request, it checks the validity of the $DCert_{SN}$ for $m_{th}$ SN. Thereafter, it is updated by the C&C center, and after both approvals, it is added to the available SN list.

**Receiver Node Registration:** All of the IoT devices in the IoT-HS that are responsible for receiving data (RNs) like doctors, paramedics, pharmacies, etc., generate private ($prk_{RN}$) and public ($pbk_{RN}$) keys. They are registered to the Blockchain C&C center by the signatory authorities RRA and C&C center by providing all the values to construct *HB*, $pbk_{RN}$ and digital certificate ($DCert_{RN}$) to the RRA, namely,

$RN_m \rightarrow RRA : requests_{RN_m} = \{HB||pbk_{RN}||DCert_{RN}\}$

As soon as the RRA receives the request, SRA checks the validity of the $DCert_{RN}$ for $RN_m$. Thereafter, it is updated by the C&C center, and after both approvals, it is added to their available list.

**Blockchain-based Protocol:**

- *Step 1:* All of the SNs periodically publish their $\tau$ to the SRA. This $\tau$ information calls for the *HB* function and updates their geo-location and operational cost (if changed due to conditions like queuing delay, low energy, etc.). It also publishes the residual energy, namely,
  $SN_m \rightarrow SRA : requests_i = \{HB||Opcost_{SN}\}$

  Every published data at time $t + 1$ replaces the $\tau$ information at t. Therefore, the SRA will replace the information of $SN_m$ at time $t + 1$ with the newly published data at $SN_m$.

- *Step 2:* All of the RNs that are beyond the IoT-HS communication periodically publish their $\tau$ to the RRA. This $\tau$ information calls for the *HB* function and updates their geo-location and operational cost (if changed due to conditions like queuing delay, low energy, etc.). It also publishes the queuing delay (QD) it is facing and urgency listing (UL) that it needs to update for SN, namely,
  $RN_m \rightarrow RRA : requests_i = \{HB||QD_{RN}||UL_{RN}\}$

  Every published data at time $t + 1$ replaces the $\tau$ information at t. Therefore, the RRA replaces the information of $RN_m$ at time $t+1$ by the newly published data at $SN_m$.

*2) Holochain Network Definition:* The Holochain technology based on the rules defined in DNA, which is a rule book for holochain, is subdivided into zomes as follows:

1. *Sender Nodes Zome (SNZ):* It is subdivided into multiple sections based on specific characteristics and monitoring requirements SNs have. These sections utilize the holochain attributes, defined below, differently.

- *ID:* The ID associated differs as per the characteristics of the section nodes. For example, Old Patients, High-Risk Patients, Regular Monitored Patients, and Athletes sections are $ODP_i$, $HRP_i$, $RMP_i$, and $AT_i$ respectively, where $i$ denotes their respective ID number.
- *Urgency:* In every section, the nodes have different urgency requirements based on their criticality, that will be generated for $ODP_i$, $HRP_i$, $RMP_i$, and $AT_i$ as 3 (highest urgency), 2, 1, and O (lowest urgency), respectively.
  $DHT(IoT - HS - SNZ) \Leftarrow RN_m \rightarrow RRA : requests_i = RRA \rightarrow C\&C : requests_i = update\{Urgency \rightarrow SN_m.ID\}$

  $DHT(IoT - HS - SNZ) \Leftarrow SN_m \rightarrow SRA : requests_i = \{HB||Opcost_{SN} ||Urgency\}$

- *Association:* This field keeps the records of which section the SN belongs to. In case of an SN urgency, then the section it belongs to is updated as follows:

  $SN_m \rightarrow DHT(IoT - HS - SNZ) : requests_i = update\{section(SN_m.ID)\}$

2. *Receiver Nodes Zome (RNZ):* It is responsible for fetching the requests and the reception of data. The DHT of the RNZ is separated from the SNZ in order to provide integrity to the data of SNs. RNZ is subdivided into sections based on specific characteristics RNs have, like Doctors, Paramedics, and Pharmacy.

- **Doctors:** This section of the RNZ requests or receives data at any time instance. The data is provided after checking its integrity to its zome. The data becomes part of a temporary allocated memory of size $TM_{f)\ddagger|}$ and then is made part of the RNZ's DHT, as follows:

    $DHT(IoT - HS - RNZ) \Leftarrow doc_m \rightarrow RRA : requests_i = RRA \rightarrow C\&C : requests_i = \{SN_m\}$

- **Paramedics:** This section of the RNZ is responsible for fetching data that has been forwarded by the RN (doctor only) or by the SNZ. No request from this section can be generated. Data received from the doctor is expected by:

    $DHT(IoT - HS - RNZ) \Leftarrow par_m : action_i = doc_m(\{SN_m\})$

    The data forwarded directly from the SNZ is first made part of DHT, and then the *action* function is called:
    $DHT(IoT - HS - RNZ) \Leftarrow par_m \rightarrow RRA : requests_i = RRA \rightarrow C\&C : requests_i = \{SN_m\}$

    $DHT(IoT - HS - RNZ) \Leftarrow par_m : action_i = \{SN_m\}$

- **Pharmacy:** This section of the RNZ is responsible for fetching data that has been forwarded to the RN (doctors/paramedics) or directly from the SNZ. Data received from doctors/paramedics is expected by:

    $DHT(IoT - HS - RNZ) \Leftarrow pha_m : action_i = doc_m(\{SN_m\})$

    $DHT(IoT - HS - RNZ) \Leftarrow pha_m : action_i = par_m(\{SN_m\})$

    The data forwarded directly from the SNZ is first made part of DHT, and then the *action* function is called:

    $DHT(IoT - HS - RNZ) \Leftarrow pha_m \rightarrow RRA : requests_i = RRA \rightarrow C\&C : requests_i = \{SN_m\}$
    $DHT(IoT - HS - RNZ) \Leftarrow pha_m : action_i = \{SN_m\}$

    **Holochain-based protocol:**
- *Step 1:* If SN's data in SNZ needs to be transferred to the RNZ, it will be shared once during a time instance $T$. If any RN wants to extract it afterward, it can request a copy from the nearest section of the RNZ in which it is available and within the IoT-HS communication.
- *Step 2:* The data fetched by RN from SN will only be updated by SNZ. Other RNs that have already requested a copy of SN's data in step 1 will get updated regularly.
    $Repeat(timeinterval) \Rightarrow DHT(IoT - HS) = doc_m \rightarrow RRA : requests_i = RRA \rightarrow C\&C : requests_i = \{SN_m\}$

### C. Time Complexity Analysis

In this subsection, we present the time complexity analysis of the proposed hybrid framework compared to Blockchain-only and Holochain-only approaches. The worst time complexity of the Blockchain-only and the Holochain-only networks is given by [20], and [21]. In the analysis provided below, the total number of validating nodes is $m$, while $c$ denotes the complexity parameter.

- **Blockchain:** Let $\Omega_{blockchain}$ be the system's complexity with $\gamma$ transactions. For every new transaction generated by the system, every node needs to check the details about the transaction and also needs to check against any double-spending by the node, resulting in time complexity as given by [21]

$$c + \gamma \quad (8)$$

In terms of Big-O notation, the $\Omega_{Blockchain}$ for each node against a transaction becomes

$$\Omega_{blockchain} \in O(\gamma^2) \quad (9)$$

Since every node has to perform a similar validation function, the time complexity thus increases by the total number of nodes $m$ in the network [21]

$$\Omega_{blockchain} \in O(\gamma^2 m) \quad (10)$$

By (10), it is clear that the system time complexity being a quadratic function makes it a bottleneck for its operation in a real-time environment, where data delivery in terms of time is critical.

- **Holochain:** Similar to the Blockchain network, Holochain has been considered to have total nodes of $m$, and its system time complexity is $\Omega_{holochain}$. Since the data transfer in Holochain is done using DHT to find a node that holds the data, a binary search at DHT is performed, and thus the time complexity is given by [20] and [21]

$$DHT\ Lookup = c + log(\gamma) \quad (11)$$

When the state transition about data is received, the node starts to gossip with its neighbors, and thus data starts to propagate. When the data arrives at an RN, the data is validated again using the DHT. The time complexity for the validation function is given by [21]

$$validation = v(\gamma, m) \quad (12)$$

Thus, combining both (11) and (12), the system-wide transition complexity is determined as

$$SysComplexity = DHTLookup + validation + Delay \quad (13)$$

Where *Delay* refers to the wireless propagation delay and the transaction processing delay. Therefore, the system-wide transition complexity is as follows:

$$SysComplexity = c + log(\gamma) + v(\gamma, m) + \Delta \quad (14)$$

In terms of big-O notation, the following applies:

$$\Omega_{holochain} \in O(n \cdot log(\gamma) + v(\gamma, m) + \Delta) \quad (15)$$

The time complexity for the holochain-only network in IoT-HS is clearly better than that of the Blockchain-only network.

- **HOLO-BLOCK CHAIN:** In our proposed hybrid Holo-Block chain framework, the authentication and registration part is only associated with the Blockchain, while the data distribution is done using holochain.

  Thus, for the authentication and registration process, the time complexity depends on the number of authorized nodes, which in our case are C&C center, RRA, and SRA. Therefore, the time complexity for the Blockchain part is lower down as follows:

  $$c + \gamma \quad \text{where } \gamma \in z \tag{16}$$

  Here $z$ is the factor that describes the transaction count allowed as per the consensus algorithm.

  By examining the Holochain part, where data transition is done using DHT, the time complexity remains the same as the Holochain-only approach and can be written as

  $$c + log(\gamma) \tag{17}$$

  Thus, including the data validation process, the combined time complexity of the proposed hybrid Holo-Block chain framework becomes

  $$(c + z) * (\gamma + log(\gamma) + v(\gamma, m) + \Delta) \tag{18}$$

  In terms of Big-O notation, the following applies:

  $$\Omega_{holo-block} \in O(z * \gamma \cdot log(\gamma) + v(n, \gamma) + \Delta) \tag{19}$$

  Although our hybrid framework's obtained time complexity is higher than the holochain-only approach, the proposed framework aids in mitigating a number of security vulnerabilities in a less complex way, as demonstrated in Section II-D.

### D. Security Analysis

The security of wirelessly communicating node's objects is always one of the main concerns. There are various security concerns in the IoT-HS, which need to be mitigated with appropriate schemes. The security measures devised in this paper against representative security concerns are as follows:

- **Impersonation Attack:** In the proposed hybrid Holo-Block chain framework, if a malicious node tries to impersonate a legitimate SN, it will not have the private key $prk_{SN}$ that is registered. Without having authorization and a digital signature from the C&C center and the SRA, it cannot interact with the network. On the other hand, in the case of a holochain-only approach [20], when the network is huge in terms of the number of nodes, the DHT also grows. So, moving through the DHT takes time, and in the meantime, the networks may get flooded by the impersonation attack.
- **Tampered Attack:** A new addition of SN in the SNZ could be done by system tampering. In the case of a holochain-only approach, [20], which is an offline chain, there is a high probability of tempering the registered list by the intruders. This cannot happen in our proposed hybrid Holo-Block chain framework due to Blockchain's tamper-proof feature that it is an online process. Thus, our proposed method saves the system from tampering attacks.
- **DoS Attack:** Even though the SC does not allow nodes that are not part of the system to execute any transaction, we could have the case where nodes become part and later on get compromised due to any reason; thus, can hit the system with a DoS attack. For such attacks, we propose an empirical method of passive blocking as follows: In the case, a node crosses a defined threshold of transfer data rate over the channel $C$, then this node is allowed to do so for a certain time period. Let's assume that the time period '**v**' is the time the node will transfer above the threshold limit, after which it will be blocked for '**u**' time intervals. Just right after blocking, it will be released again. If it again crosses the threshold limit for another '**v/2**' time period, it will be punished and put to blockage exactly double the time period **2u** to which it was previously blocked. After that, it will again be monitored, and if the behavior is similar for the '**v/4**' time interval, it is permanently put into the blocking list. In contrast, a holochain-only approach provides no way to mitigate such a DoS attack proactively so as to stop the network from getting poisoned. This proposed method of passive blocking is evaluated in Section III-B.
- **Man-in-the-Middle (MiM) Attack:** In [20], a holochain-only approach takes the hash of the local chain and puts it alongside the packet so that no one can change it. However, the authorization of a legitimate user is done using the DHT, and the node has to take the packet in order to check its authenticity. In our proposed hybrid approach, only the legitimate user can interact with the system and can then take advantage of the holochain feature of attaching the hash to the packet, along with copying it to the DHT.

## III. Performance Analysis

In this section, we evaluate the proposed Holo-Block chain framework in IoT-HS in terms of performance and network liability. Firstly, we provide the time complexity analysis, the memory, and computational cycles cost estimation for the proposed hybrid framework compared to the Blockchain-only and Holochain-only [20] approaches. We then perform network analysis for the proposed hybrid framework and compare it with its counterparts in terms of the network latency, the normalized average network throughput, and route selection time. Finally, we analyze the proposed hybrid system response against the DoS attack.

### A. System Configuration

The experiments are carried out on Intel $Core^{TM}$ i7-10700 CPU 2.90GHz (16CPU) Processor with 16 GB RAM. We have utilized an IoT-HS prototype in MATLAB to validate the feasibility and effectiveness of the proposed protocol. For the proposed Holo-Block chain framework, we were able to construct a distributed application using holochain and blockchain functionality through JavaScript and Go language implementation (Geth) along with web3py [22], respectively. For the inter-connectivity, we took advantage of the socket concept, and MATLAB could interact with Holochain and Ethereum network.

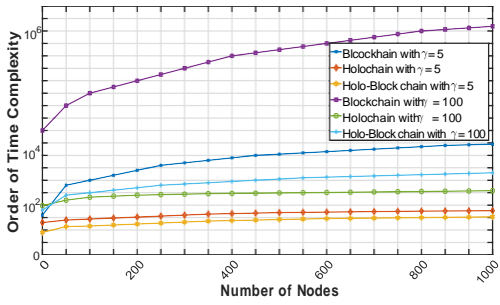

Fig. 3: Time Complexity

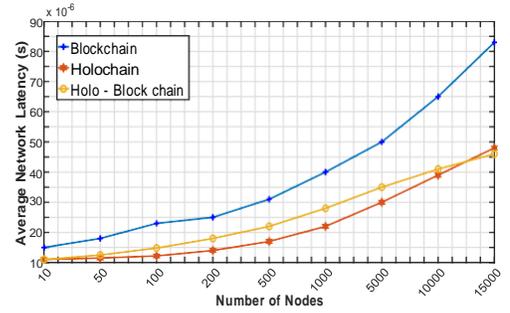

Fig. 5: Network Latency ($\gamma = 100$)

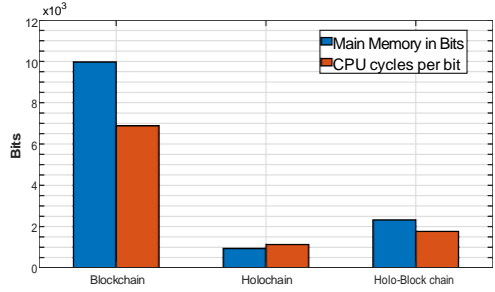

Fig. 4: CPU Usage & Memory Utilization ($\gamma = 100$)

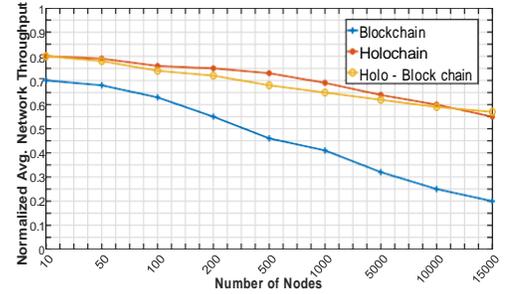

Fig. 6: Normalized Network Throughput ($\gamma = 100$)

*B. Results*

**Time Complexity:** As discussed in Sections II-C and II-D, it is clear that Blockchain will always have high time complexity due to the need for authentication being carried out for each transaction. Also, as holochain [20] looks for a transaction only from the DHT, it has the least time complexity. In contrast, in the case of the proposed hybrid Holo-Block chain framework, the time complexity gets a little high compared to the holochain-only approach due to the additional authentication layer over holochain. Fig. 3 clearly depicts the trends where we performed time complexity calculations for 5 and 100 transactions for each of the schemes.

**Main Memory and CPU Cycles:** In Fig. 4, it can be observed the main memory utilization and required computational power. It is clearly seen that Blockchain has high demands for main memory because of its replication process. In our approach, since we utilize the layer of authentication by Blockchain, this makes the main memory requirements to be higher than the holochain-only approach [20], but far lesser than the blockchain-only mechanism. This is because data dealing in zomes is controlled by Holochain.

**Network Latency:** In Fig. 5, the Blockchain implementation provides the highest network latency due to the consensus algorithm. The proposed hybrid framework manages to have the lowest network latency when the number of nodes significantly increases. This is because the authentication is handled by the blockchain part of the hybrid framework, which never allows intruders to enter the network and interact with others. On the other hand, in the case of the holochain-only approach, the offline mechanism makes the intruder send data while others may look at the DHT for its authenticity; thus, it may severely flood the network.

**Normalized Average Network Throughput:** Fig. 6 shows the normalized average network throughput as the number of nodes increases. The blockchain-only approach has the lowest normalized network throughput because of the delay that occurred during the authentication mechanism applied for every transaction. On the other hand, the holochain-only approach has lower network throughput compared to the proposed hybrid framework when the number of nodes in the network increases significantly. This is because, in the holochain-only approach, the time for authentication through DHT gets higher while an intruder can flood the network with malicious packets. In contrast, the proposed hybrid framework never allows any intruder to even interact with the network.

**Route Selection Time:** When a node needs a secure path to route a packet, the Blockchain is the best policy. However, route selection time may increase since every route needs to be registered on Blockchain. In contrast, our approach is not registering every route, and no one can act on behalf of the registered node, thus improving the overall route selection time as shown in Fig. 7.

**DoS Attack:** In Section II-D we have proposed an approach to mitigate DoS attacks launched by any of the registered nodes that get compromised due to various reasons, i.e., environmental changes, terrain uncertainty, malfunctioning, etc. In Fig. 8, it can be seen how the attack can be mitigated. The threshold of the transfer data rate is set to 0.9 Gbps, and the initial allowed time is v=10min and is dropped to 5min while the blockage time is increased from u=5min to 10min.

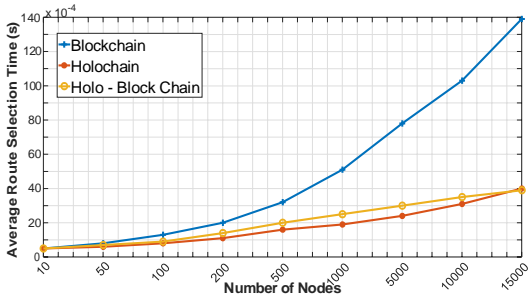

Fig. 7: Route Selection Time ($\gamma = 100$)

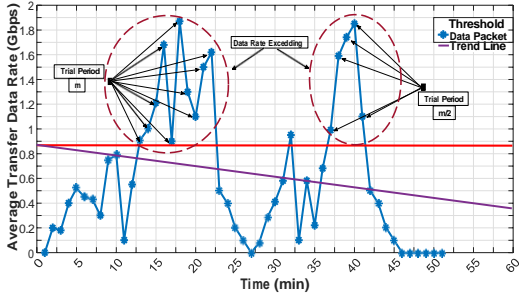

Fig. 8: DoS Attack Mitigation

## IV. Conclusions and Future Work

A hybrid Holo-Block chain framework for secure communication in the Internet of Things Healthcare Ecosystem (IoT-HS) has been proposed that enables a low computational, low in memory, fast, secure, and reliable network. In contrast to a blockchain-only approach, the proposed hybrid approach has the advantage of less computational and memory consumption requirements. Also, compared to a holochain-only approach, the proposed approach enables the authentication mechanism to be fully distributed so that the network is not poisoned with attackers' packets. Through thorough performance analysis, it is demonstrated that the proposed hybrid Holo-Block chain framework provides better reduction in time and space complexity as compared to the blockchain-only scheme but slightly higher compared to the holochain-only one. Furthermore, the proposed hybrid framework provides an advantage over its counterparts in terms of network performance, i.e., network latency, network throughput and route selection time, and attacks mitigation. Therefore, the hybrid framework is demonstrated to be an efficient solution for IoT-HS. In future work, we will investigate the automation of identifying malicious activity by the internal nodes of the hybrid Holo-Block chain framework using Deep Learning techniques.